\documentclass[aip,jcp,twocolumn,10pt]{revtex4-1}
\pdfoutput=1
\usepackage{color} 

\definecolor{darkblue}{rgb}{0,0,0.5}
\definecolor{lila}{rgb}{0.3,0,0.3}
\definecolor{turq}{rgb}{0,0.1,0.4}
\definecolor{lightblue}{rgb}{0.7,0.7,0.9}

\usepackage[pdftex,
  colorlinks=true,
  backref=page,
  linkcolor=darkblue, 
  filecolor=red,
  citecolor=turq, 
  urlcolor=lila, 
  pdftitle={Excitation and emission spectra of rubidium in rare-gas thin-films},
  pdfauthor={Ilja Gerhardt, Brian Sin, Takamasa Momose},
  pdfsubject={Excitation and emission spectra of rubidium in rare-gas thin-films},
  pdfkeywords={Rubidium, rare-gases, matrix isolation, excitation spectra, atomic spectroscopy},
  pdfpagelabels=true, 
  breaklinks=false,
  plainpages=false,
  backref=false,
  bookmarks,
  bookmarksnumbered=true]{hyperref}

\usepackage[table]{xcolor}
\usepackage{tabularx}

\usepackage{graphicx,epstopdf}

\newcommand{\nozzlesample}{33} 
\newcommand{\dispensersample}{40} 
\newcommand{\arthickness}{2.7} 
\newcommand{\krthickness}{3.2} 
\newcommand{\xethickness}{4.1} 
\newcommand{\arem}{877} 
\newcommand{\aremwidth}{53} 
\newcommand{\kriex}{722} 
\newcommand{\kriiex}{731} 
\newcommand{\kriiiex}{743} 
\newcommand{\krem}{923} 
\newcommand{\kremwidth}{37} 
\newcommand{\xeiex}{760} 
\newcommand{\xeiiex}{771} 
\newcommand{\xeiiiex}{782} 
\newcommand{\xeem}{989} 
\newcommand{\xeemwidth}{39} 

\begin{document}

\title{\large{Excitation and emission spectra of rubidium in rare-gas thin-films}}

\author{Ilja Gerhardt}
\email{ilja@quantumlah.org}
\altaffiliation{Current address: Max Planck Institute for Solid State Research, Heisenbergstra\ss e 1, D-70569 Stuttgart, Germany}
\affiliation{The Department of Chemistry, Low Temperature Group, The University of British Columbia, Vancouver, B.C.\ V6T~1Z1, Canada}

\author{Brian Sin}
\affiliation{The Department of Chemistry, Low Temperature Group, The University of British Columbia, Vancouver, B.C.\ V6T~1Z1, Canada}

\author{Takamasa Momose}
\affiliation{The Department of Chemistry, Low Temperature Group, The University of British Columbia, Vancouver, B.C.\ V6T~1Z1, Canada}


\begin{abstract}
To understand the optical properties of atoms in solid state matrices, the absorption, excitation and emission spectra of rubidium doped thin-films of argon, krypton and xenon  were investigated in detail. A two-dimensional spectral analysis extends earlier reports on the excitation and emission properties of rubidium in rare-gas hosts. We found that the doped crystals of krypton and xenon exhibit a simple absorption-emission relation, whereas rubidium in argon showed  more complicated spectral structures. Our sample preparation employed in the present work yielded different results for the Ar crystal, but our peak positions were consistent with the prediction based on the linear extrapolation of Xe and Kr data.  We also observed a bleaching behavior in rubidium excitation spectra, which suggests a population transfer from one to another spectral feature due to hole-burning. The observed optical response implies that rubidium in rare-gas thin-films is detectable with extremely high sensitivity, possibly down to a single atom level, in low concentration samples.
\end{abstract}

\maketitle

\section{Introduction}
The strong $S \rightarrow P$ transitions of alkali atoms  have attracted a large interest in spectroscopy for a long time~\cite{herzberg_book}. Gas phase experiments of these alkaline systems have led to outstanding findings in quantum optics and physics. Whereas gas phase spectra can be fully theoretically described, spectra of alkali atoms in solid states depend on environmental influences, which have not been understood well. Metal atoms embedded in rare-gas hosts were studied sparsely in the past decades~\cite{meyer_tjocp_1965,balling_jcp_1978,tam_tjocp_1993,crepin-gilbert_iripc_1999,moroshkin_pr_2008,ryan_tjopca_2010,xu_prl_2011}. These studies are useful for characterization of the guest atom, its local environments, and its transition properties. It further gives insight into the trapping sites. The Jahn-Teller effect can be characterized, since the electronic transition is modified by the highly sterical distorted lattice.

For quantum optics applications, a highly sensitive fluorescence detection of low-doping samples  is desired. So far only a few studies characterize the \emph{emission} properties of various metal species in different host systems. If the optical properties of alkali atoms are well understood, it might also give a better understanding of the optical desorption of atoms from surfaces~\cite{klempt_pra_2006}.

Atomic rubidium (Rb) in its vapor form is widely used in experimental quantum and atom optics. It is convenient because of its laser wavelengths and the spectral separation of the two $D-$line transitions. Comparatively little is known about optical properties of Rb embedded in solid state matrices. Early experiments on the absorption spectra of Rb in argon were performed in the 1960s by \textsc{Kupfermann} and coworkers~\cite{kupferman_pr_1968}. But these comprehensive studies were limited to spectra in argon crystals. Only a short qualitative report exists on the luminescent properties of the sample~\cite{balling_jcp_1983}. Experimental studies of Rb in solid helium (He) observed weak emission properties~\cite{eichler_prl_2002,hofer_pra_2006}. In liquid He, Rb was reported to show strong  laser induced fluorescence~\cite{kinoshita_prb_1994,kinoshita_pra_1995}. Further investigations of Rb doped He nanodroplets~\cite{theisen_epjd_2011} showed that laser induced fluorescence can be observed in this system, too. A sample produced by $\beta^{-}$ decay of krypton, resulting in Rb dopant atoms was researched by \textsc{Micklitz} and \textsc{Lucher}~\cite{micklitz_zfpahan_1974}. The spectral emission was found to depend strongly on their sample preparation method. In all the above studies, the reported optical characterizations focus on the $D-$line transitions in the near-infrared part of the optical spectrum. These transitions are often blue shifted from the corresponding gas phase transitions and are spectrally broader by serveral orders of magnitude.

In this paper we report on the absorption and emission properties of Rb in various rare-gas thin-films. We prepared `spray-on' crystals of argon (Ar), krypton (Kr), and xenon (Xe)  doped with rubidium from a thermal dispenser.  The thermal dispenser has been commonly used in quantum optical experiments with atoms; we are not aware of any other paper which describes a similar sample-preparation method for doped thin-film measurements.  Laser ablation or electronic discharge has been often used for sample preparation in matrix isolation studies of atoms. In this study, excitation-emission two-dimensional spectra were recorded in order to allow more insight on the properties of Rb in a solid state environment. We failed to acquire spectra with Rb in neon (Ne), although the cryogenic conditions allow for the growth of doped Ne crystals. Likewise, experiments with nitrogen (N$_2$) yielded a null result and were not  pursued further.

\begin{figure}[b]
\centering
\includegraphics[width=8.5cm]{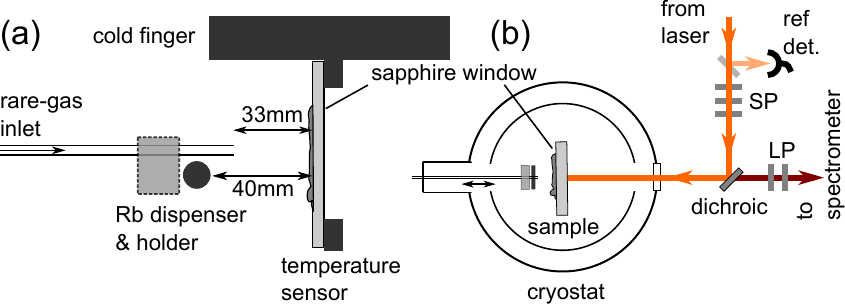}
\caption{Experimental setup used in the present study. (a) Side-view inside the cryostat. A nozzle, fed from a compartment at ambient temperature resides inside the vacuum jacket of the cryostat. The thermal rubidium dispenser attached on a thermal protective holder is held below the nozzle. The entire nozzle-dispenser arrangement can be mechanically retracted inside the vacuum chamber and replaced with a black beam block for excitation/emission measurements. (b) Top-view of the optical setup. An optical short-pass filter (SP) reduces unwanted higher order contributions from the laser and background fluorescence of the optical fiber leading to the experiment. Two long-pass filters (LP) allow a highly sensitive detection of Stokes-shifted emission from the doped thin-films. The sapphire window is slightly tilted to reduce the scattering and reflection of laser light to the detection channel.\label{fig:setup}}
\end{figure}

\section{Experimental Details}
The rare-gas thin films were grown  on  a 1~mm thick sapphire window thermally connected to the cold finger of a closed cycle GM refrigerator (Sumitomo, RDK-408).  The base  temperature of the cryostat was 4.1~K and  its cooling power was 1~W. Rare gases were dosed from a nozzle (inner diameter=1.7~mm) mounted  \nozzlesample~mm in front of the window. A Rb-dispenser (Alvatec, AS-3-Rb-25-S) held on a thermal protective plastic holder (ULTEM) was attached at the tip of the nozzle, and Rb vapours generated by the dispenser were co-deposited on the sapphire window during the thin film growth. Earlier experiments with a chromate based Rb dispenser (SAES Getters, 5G0120) showed broad spectral features due to impurities. 

The rare-gases were supplied through a flow-controller (Horiba STEC 7320) with a flow rate of 50~ccm (cubic centimetre per minutes). The gas was expanded from ambient conditions and both the nozzle and the dispenser holder were kept at ambient temperatures. After the deposition, the nozzle/ dispenser arrangement was retracted  $\approx$ 300~mm away from the crystal window, and obscured by an opaque matte block to prevent the residual fluorescence of the high-temperature plastic holder from reaching our detector. Prior to each experiment, the residual fluorescence of the cleaned sapphire window was verified to be at the noise-level for acquisition times of 60~secs. 

For optical absorption measurements, a retractable mirror was installed next to the crystal window to steer the optical beam through the sample. Optical absorption spectra were acquired with a tungsten lamp and a double-grating monochromator (Spex). The instrumental linewidth was measured to be below 1.3~nm.  The light dispersed by the monochrometer was detected by a photon counting photomultiplier (Hamamatsu) through an optical multimode fiber. 

A continuous wave (cw) titanium-sapphire (TiSa) ring laser was used (Coherent, 899-21, SW optics) for excitation and emission measurements of Rb / rare-gas samples. The optical etalon unit of the ring laser was removed to achieve a wide continuous tuning range between 690 and 790~nm. The laser line width was narrower than 0.3~nm without the etalon. The light from the ring laser was directed to the cryostat via an optical single mode fiber. The incident light was linearly polarized and its maximal intensity was measured to be $\approx$ 200~mW over a Gaussian beam profile with 2~mm diameter. A calibrated photo diode was used to monitor the incident power throughout the experiment. A combination of three short-pass filters (one SP785 `RazorEdge', Semrock, NY and two FES0800, Thorlabs, NJ) removed any incident radiation longer than  785~nm from the laser or the optical setup. The excitation light was reflected into the cryostat by a dichroic mirror (FF801-Di02, Semrock). Emission from the sample longer than 800~nm passed through the same dichroic mirror, followed by two long pass filters (LP780 and LP785, Semrock), and  was collected into a 50~$\mu$m multimode fiber (Thorlabs).  The long pass optical filters did not exhibit strong distortions to the acquired emission spectra above $\lambda$=810~nm. To further reduce the amount of reflected and scattered light going into the detection path, the sample sapphire window was tilted by a few degrees from the optical axis. Emission spectra were spectrally analyzed by a spectrometer with a 300 lpi grating (Acton SP-2356 Imaging Spectrograph), and detected by a sensitive CCD camera (Roper Scientific, PI Acton PIXIS:100B).

Special attention was given to ensure reliable and reproducible sample preparation. The following procedure was programmed and computer-controlled: (1) After achieving the base temperature of 4.2~K for more than 20~minutes, pure rare-gas films were first grown on the sapphire window for 300~s with 50~ccm gas flow,  (2) then the doped crystal of rare-gas and rubidium was grown for 1500~s with the rare-gas flow held constant, and (3) finally pure rare-gas films were coated on the mixed crystal by dosing pure rare-gas for another 300~s. The films were grown at 7.8~K, which gave better crystallinity. The sample holder temperature did not rise during the crystal growing process. Annealing of the films for several hours at T$\approx$T$_{\rm melt}/3$ was also attempted, but did not produce any qualitative change in the acquired spectra in any matrices. All the spectral observation was made at 4.2~K. 

In order to achieve a high concentration of isolated atoms and not clusters, the dispenser was tuned to a high current (8~A) during the growing process. This corresponds to a measured temperature of the dispenser of 250-300$^{\circ}$C. The dispenser had a small and narrow opening (0.5$\times$4~mm$^2$), and was positioned \dispensersample~mm from the sapphire window. The concentration of dopant atoms inside the crystals was estimated to be on the order of 1:1000. 
We found that an initial heating of the dispenser was required to achieve a reproducible and homogeneous rubidium sample. We annealed the dispenser far away from the sapphire window for 20~min prior to each crystal growth.

\section{Results}
\subsection{Film Thickness}
An interferometric measurement was used for the measurement of the thickness of thin-films. Single-frequency HeNe-Laser radiation (632.8~nm) was focused to a small point ($\approx$10~$\mu$m), centered on the sapphire window, and the optical fringes after the thin-films was recorded during crystal growth. The spacing of the observed oscillations depended linearly on growing time. Since the thickness measurement was checked against substantial thicker samples, the bulk refractive indices taken from literature~\cite{smith_pm_1961,sinnock_pr_1969,sinnock_jopcssp_1980} were used for the thickness calculation. The typical thicknesses for the doped fraction of the Ar, Kr, and Xe films, were \arthickness, \krthickness, and \xethickness ~$\mu$m respectively. The simultaneous doping of Rubidium did not affect the final thickness of the film; the observed optical interference fringes showed the same spacing with and without current flow in the dispenser. With the above mentioned flow rates and times, we found that $\approx$0.1\% of the sprayed on gas adsorbs on the window. This value was consistent for Ar, Kr and Xe. The low yield indicates a continuous ablation from the film while growing, and explains the lack of differences between annealed and unanealed films.

\begin{figure}[h!t]
\centering
\includegraphics[width=8.5cm]{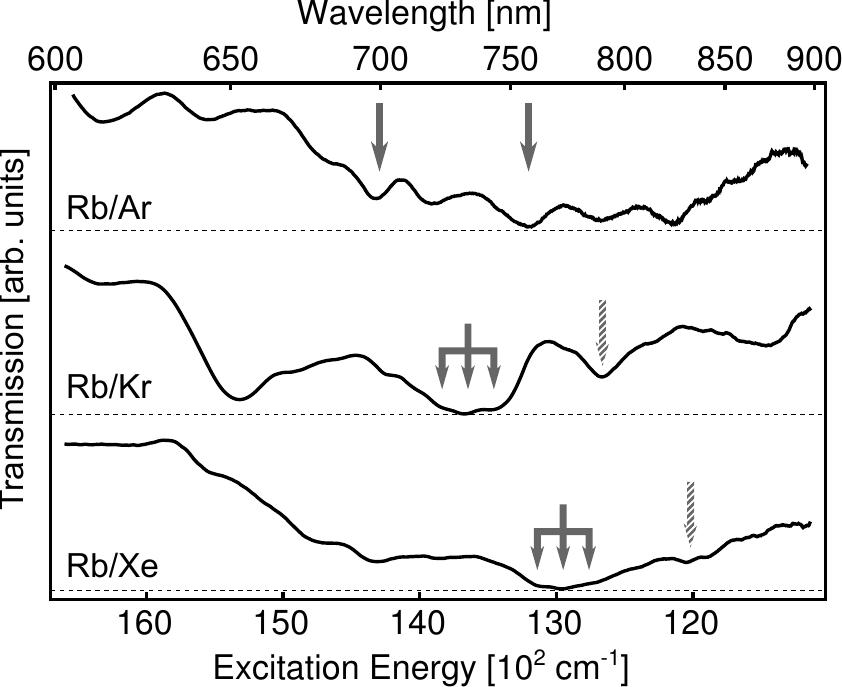}
\caption{Absorption spectra of rare-gas thin films doped with rubidium.  The arrows indicates the observed excitation maxima (Fig.~\ref{fig:2d}). For argon we observe a repeating structure with strong peaks at 705 and 755~nm. The Rb/Kr and Rb/Xe triplets are visible in the excitation scans and show small dents in the currently observed transmission spectra. The predicted positions of the red triplets for the Rb/Kr and the Rb/Xe sample are shown as dashed arrows.\label{fig:abs}}
\end{figure}

\subsection{Absorption Spectra}
Optical absorption spectra of Ar, Kr and Xe thin-films doped with rubidium are shown in Fig.~\ref{fig:abs}. A good qualitative agreement with previously reported spectra~\cite{kupferman_pr_1968} was obtained for the Rb/Ar system. This reconfirms the existence of two distinct triplets, originating from the removal of the threefold orbital degeneracy of the {\it P} configuration due to the Jahn-Teller crystal field effect. The triplets for the Rb/Kr and the Rb/Xe systems were observed as small dents in the acquired spectra. 

\begin{figure*}[h!t]
\centering
\includegraphics[width=16cm]{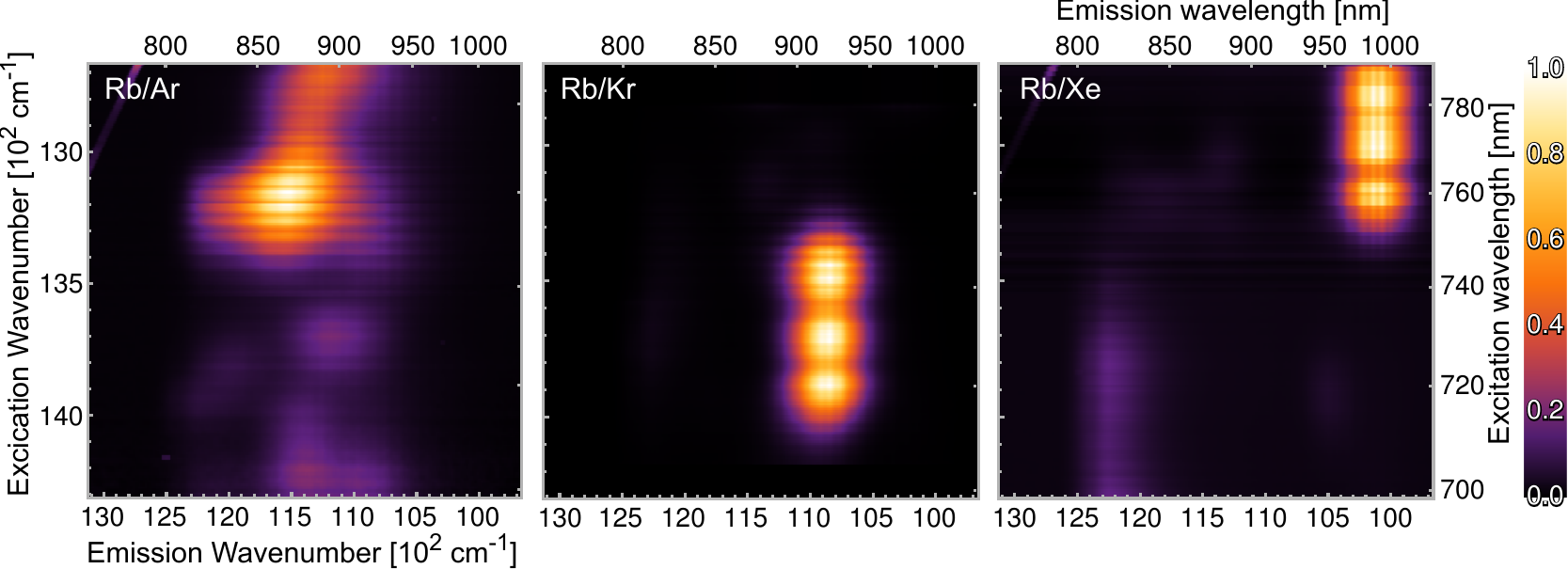}
\caption{Two-dimensional normalized excitation-emission spectra of rubidium in argon, krypton and xenon, observed at T=4.2~K. The argon spectrum shows two main contributions, at the excitation wavelength around 705~nm and 755~nm. These values correspond roughly to the earlier reported values for the red and blue triplet~\cite{kupferman_pr_1968}.  The commonly observed simple triplet structure for alkali atoms in rare-gas hosts is seen in Kr and Xe systems. Emission from Rb/Kr was around one order of magnitude stronger than that  for the Rb/Xe system.\label{fig:2d}}
\end{figure*}

\subsection{2D Spectrum in Ar}
Since the emission properties of Rb in rare-gas hosts are not well characterized in the literature~\cite{balling_jcp_1983},  excitation-emission two dimensional spectra were recorded for the full characterisation. The emission spectra were recorded at a low irradiance on the order of 5 kW/m$^2$.

The two dimensional spectra of Rb in Ar are shown in the left panel of Fig.~\ref{fig:2d}. A number of emission-excitation peaks were observed. The same complicated 2D spectrum was observed even after the purity of the used argon was carefully checked. After additional annealing for several hours, neither the qualitative nature, the peak positions, nor their relative intensity of the spectrum did change. Argon crystals were prepared with several different conditions, but the observed spectra were consistent over a large range of preparation conditions. 

In the excitation spectrum, we found peaks around 705~nm and 755~nm, which are also seen in Fig.~\ref{fig:abs}. The separation of these two peaks  in our spectrum is smaller than that reported earlier (700~nm and 775~nm~\cite{kupferman_pr_1968,balling_jcp_1983}). The peak in emission due to the excitation of the red-triplet occurred at around \arem ~nm, which is higher than the reported value of 830~nm~\cite{balling_jcp_1983}. Generally, the observed emission in Ar was weaker than in other rare-gas hosts. The difference between our results and the previous reports may be attributed to the difference in crystal growth conditions. Due to the lattice mismatch between Ar and Rb, the emission spectrum must be very sensitive to local lattice structures.

\begin{figure}[h!b]
\centering
\includegraphics[width=8.5cm]{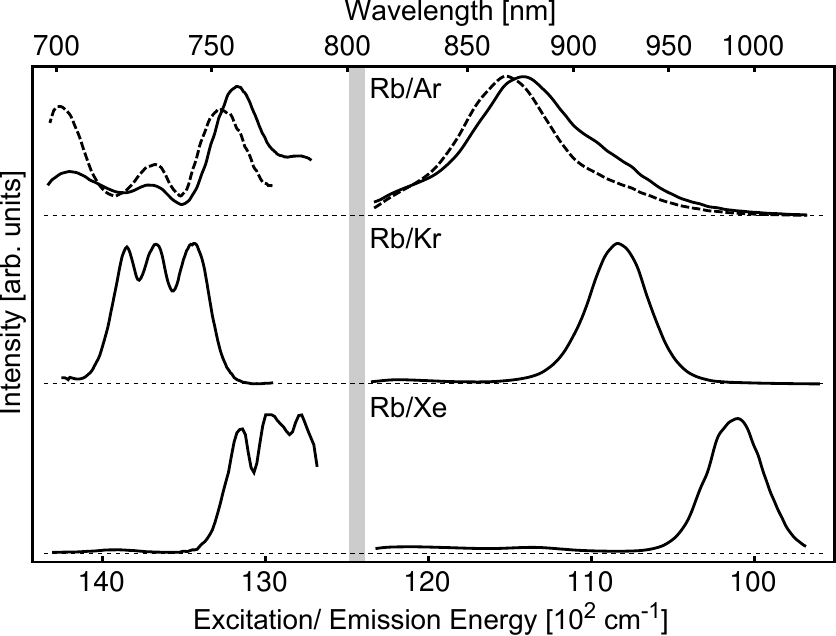}
\caption{Excitation (left) and emission (right) spectra of Rb in rare-gas thin-films obtained by integrating 2D spectrum in Fig.~\ref{fig:2d} along the emission and excitation axes, respectively. The dashed line in Ar shows the spectra obtained by integrating a limited range, centered around the strongest emission peak at 755~nm for excitation and 870~nm for emission. \label{fig:ex}}
\end{figure}

Fig.~\ref{fig:ex} shows excitation and emission spectra, which were obtained by integrating 2D spectra shown in Fig.~\ref{fig:2d}.  Direct comparison with the absorption spectrum in Ar (see the top trace of Fig.~\ref{fig:abs}, note that the absorption peaks are downwards) and the excitation spectra (see the upper right panel of Fig.~\ref{fig:ex}, note that the excitation peaks are upwards) tells us that the emission in the Rb/Ar system is mostly due to the very efficient absorption around 755~nm.

In a number of experiments we realized that the amount of laser induced fluorescence was highest for previously-unexposed samples. Further investigations on the Rb/Ar system indicated a bleaching behavior at very high laser irradiation ($I\approx$10~MW/m$^2$).  Specifically, the irradiation of intense laser radiation at 710~nm depleted the excitation peak at 705~nm with the increase of excitation between 730~nm and 770~nm~nm; meanwhile the irradiation at 760~nm weakened the excitation signal at 755~nm with the increase of excitation between 700~nm and 750~nm. These behavior suggested  that the bleaching is associated with the population transfer between different sites. Further study is needed to identify site specific behaviors in the bleaching.

The bleaching behaviour is described by the equation $a\,exp(-I t/\tau)$, where $a$ is a constant scaling factor, $I$ is the irradiance in Wm$^{-2}$ and $\tau$ is a characteristic time constant. The associated time-constant for the Rb/Ar system was $\tau$ = 1.2~${\rm W s m^{-2}}$. Such bleaching only occurred at high laser intensity and was not observed during the low-intensity spectroscopic measurements.
 
\subsection{2D Spectrum in Kr}
The two dimensional spectrum of rubidium in krypton (Fig.~\ref{fig:2d}, middle), exhibits a simple triplet of excitation-emission peaks. The three peaks are of comparable size and approximately equally spaced in energy. They are centered around \kriex, \kriiex, and \kriiiex~nm respectively, and can be assigned to the so-called ``blue-triplet''~\cite{kupferman_pr_1968}. The corresponding red-triplet was not detectable within our detection wavelength range. The corresponding emission spectrum is centered around \krem ~nm with \kremwidth ~nm full-width at half-maximum (FWHM).

No spectral difference between annealed and non-annealed samples were detected for Rb in Kr. The photo-depletion, however, of Rb in Kr was observed. After illuminating the samples with an irradiance of $I$=10~MW/m$^2$, we found about 95\% of the signal had disappeared. The remaining signal was very stable for further illumination. Although we did not observe any peaks increase in height after the photo-illumination, we tentatively attribute this depletion to a hole-burning process as in the case of the Rb/Ar system. The associated time-constant was $\tau$ = 1.1~${\rm W s m^{-2}}$.

\subsection{2D Spectrum in Xe}
The rubidium spectrum in xenon (Fig.~\ref{fig:2d}, right) shows a typical triplet structure as for Rb in Kr. The three peaks were close to the edge of our excitation range and the two peaks at longer wavelength were weakly resolved. The excitation wavelengths of these peaks were measured at  \xeiex, \xeiiex, and \xeiiiex~nm respectively. In addition, weak peaks were observed near 810~nm in the emission wavelength.  These weak peaks were not studied in further detail, since the clipping of spectra was severe due to the limited detection range. Generally, the quantum yield of the xenon emission was about one order of magnitude less than that of the krypton samples. We found an increase of scattering of white light for thicker Xe films. This indicates that the Xe films are not as homogeneous as other Ar and Kr films. Xenon's higher melting point explains this different crystallization behavior. 

\begin{table}\footnotesize
\label{table:matrix}\centering %
\newcolumntype{Y}{@{\extracolsep{5mm}}>{\columncolor[gray]{0.92}}c@{}}
\newcolumntype{W}{@{\extracolsep{\fill}}>{\columncolor[gray]{0.92}}c@{}}
\newcolumntype{Z}{>{\centering\arraybackslash}X}
\renewcommand{\tabularxcolumn}[1]{>{\arraybackslash}m{#1}}

\begin{tabularx}{85mm}{@{}p{1.6cm}@{}ZZZZ@{}}
\rowcolor[gray]{0.92}
{\bf Observed}&\multicolumn{4}{W}{{\bf Contribution}} \\
\rowcolor[gray]{0.92}
{\bf system}&Excitation wavelength& FWHM&Emission wavelength&FWHM \\
\rowcolor[rgb]{0.85,1,0.85}
Rb/Ar, red triplet\cellcolor[gray]{0.92}  &755&15& \arem & \aremwidth \\
\rowcolor[rgb]{0.85,1,0.85}
Rb/Ar, blue triplet\cellcolor[gray]{0.92} &705& 15 & \arem & \aremwidth \\
\rowcolor[rgb]{0.85,1,0.85}
Rb/Kr, blue triplet\cellcolor[gray]{0.92}  & \kriex, \kriiex, \kriiiex & 29 & \krem & \kremwidth \\
\rowcolor[rgb]{0.85,1,0.85}
Rb/Xe, blue triplet\cellcolor[gray]{0.92}  & \xeiex, \xeiiex, \xeiiiex & 30 & \xeem & \xeemwidth \\
\end{tabularx}
\caption{Wavelengths of the excitation and emission of each peak, and thier linewidths.}
\label{tab1}
\end{table}

Table~\ref{tab1} summarizes the observed peak wavelengths in all crystals.  

\subsection{Spectra in Ne}
We have attempted to observe Rb spectra in Ne films, but we did not detect any signal. We also tried a different sample preparation method, in which the system was subject to 1000 repeated growing-cooling cycles (1~sec of deposition and 3~sec of cooling).  This method also failed to produce spectra.  We attribute the failure in Ne samples to the close proximity of the nozzle and dispenser to the sapphire window under our experimental setup.  Precooling the Ne gas may improve the growth of doped Ne thin-films.

\section{Discussion}
\begin{figure}[h!t]
\centering
\includegraphics[width=8.5cm]{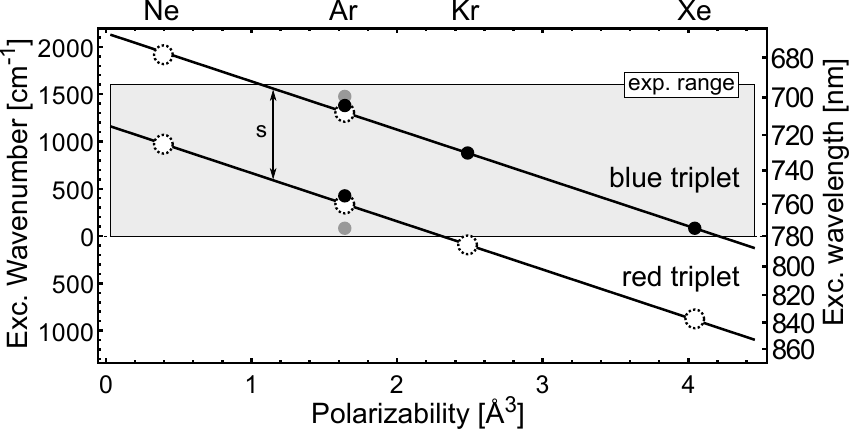}
\caption{A plot of excitation wavelength vs the polarizability of the rare-gas hosts. The {\it y}-axis is linearly scaled in energy.  The left-axis shows the wavelengths relative to the gas phase $D_2$ transition. The black points indicates the observed wavelengths obtained in the present study, and the grey points are those reported earlier~\cite{kupferman_pr_1968}.  
The dashed circles indicate extrapolated wavelength to predict the positions of undetected peaks. The extrapolation assumed the same spacing between each pair of  blue and red triplets as in Ar (denoted by $s$) \label{fig:pola}}
\end{figure}

The spectral shift of the $S \rightarrow P$ transition in a solid state system depends on the host environment. In  rare-gas films, it is expected that the atomic polarizability contributes most of the spectral shift. In Fig.~\ref{fig:pola} the spectral positions of the blue and red triplet are shown referenced to the gas phase $D_2-$line.  The close linear relation between excitation wavenumber and the polarizability allowed assignment of the red and blue bands, partially through extrapolation.  The relation between emission wavenumber and the polarizablity was also found to be linear, and  has been reported earlier~\cite{ryan_tjopca_2010}. 

By extrapolating the blue triplets of Kr and Xe linearly, we estimate the excitation wavelength in Ar to be around 708~nm as seen in Fig.~\ref{fig:pola}.  If we assume a linear relation for the emission wavenumbers as well, we estimate the emission in Ar to be  around 890~nm extrapolating from the values of  \krem \ and \xeem ~nm in Kr and Xe, respectively. These predicted excitation and emission peak wavelengths are closer to our detected peaks (705~nm for excitation and 877~nm for emission) than those reported previously (700nm for excitation and 830~nm for emission)~\cite{kupferman_pr_1968,balling_jcp_1983}.  Experiments on sodium in rare-gas hosts~\cite{ryan_tjopca_2010} showed a similar spectral deviation for the Ar system. 

By assuming the same separation between the blue and red triplets for each gas, we can estimate the positions of red triplet peaks in Kr and Xe.   As shown in Fig.~\ref{fig:pola}, the predicted excitation wavelengths are out of the range of our laser coverage. This explains why we did not observe peaks corresponding to the red triplet in Kr and Xe. 

Complicated excitation-emission 2D spectrum in Ar compared to Kr and Xe may result from size differences between the host and matrix atoms.  Since there is no report on the one-to-one pair bond length between Rb and rare gas atoms,  we employ the reported van der Waals (vdW) radii of 188~pm (Ar), 202~pm (Kr), 216~pm (Xe)~\cite{bondi_tjopc_1964,mantina_tjopca_2009}, and 303~pm (Rb)~\cite{mantina_tjopca_2009} for the size of each atom as a simple approximation. The rare-gas lattice must be distorted by the presence of Rb occupying a single substitutional site. The greater mismatch in radius for Rb/Ar leads us to expect a larger lattice distortion than that from Rb/Kr and Rb/Xe, which may explain the complicated excitation-emission 2D spectrum in Ar.  More quantitative discussion on the spectral impact of crystal defects was reported previously~\cite{micklitz_zfpahan_1974}. 

The spectral splitting between the three peaks of the observed triplets in Kr and Xe characterizes the Jahn-Teller splitting due to interactions with the host lattice. We do not find any difference between the two Jahn-Teller splittings in Kr and Xe within our measurement accuracy. The interaction that induces  the Jahn-Teller splitting must be similar in both gases

\section{Conclusion}
We have presented absorption, excitation and emission spectra of rubidium in rare-gas hosts. For the excitation and emission, we used two-dimensional spectroscopy to characterize the spectra in more detail. Certain slices through the two-dimensional Rb/Ar specrta agree with earlier reports; however, we find the emission wavelength for the Rb/Ar system red-shifted by $\approx$40~nm. The spectra of Rb in Kr and Xe have a much simpler structure; the Rb/Kr system showing the strongest emission. Therefore, for very low concentration samples approaching to the single atom level, we propose  the use of the Rb/Kr system. For such experiments we would need to increase the collection efficiency of the cryogenic confocal microscope.

Our results will be extended to lower Rb concentrations and open a way to better understand the optical properties of thin films of Rb in rare-gases. Further experiments with magnetic fields are under preparation~\cite{schrimpf_zfpbcm_1987}. The undesirable adsorption of Rb and other atoms on surfaces of the vacuum chambers is observed in a number of atom-optics experiments. The optical desorption of Rb from surfaces has been observed~\cite{klempt_pra_2006} and stands as a potentially useful mechanism. Our experiments might allow a more efficient desorption, since the optical spectrum is better understood.

\section*{Acknowledgments}
We thank Pavle Djuricanin at the Center for Research on Ultra-Cold Systems (CRUCS) for technical support. Birinder Kahlon assisted in the initial steps of preparing the cryogenic setup. I.G. acknowledges the careful proof-reading of this manuscript by Dr.~P.~R.~Hemmer and Eric Miller. This work is supported by an NSERC Discovery Grant and funds from CFI to CRUCS at UBC.\\


\begin{thebibliography}{10}%
\makeatletter
\providecommand \@ifxundefined [1]{%
 \ifx #1\undefined \expandafter \@firstoftwo
 \else \expandafter \@secondoftwo
\fi
}%
\providecommand \@ifnum [1]{%
 \ifnum #1\expandafter \@firstoftwo
 \else \expandafter \@secondoftwo
\fi
}%
\providecommand \enquote [1]{``#1''}%
\providecommand \bibnamefont  [1]{#1}%
\providecommand \bibfnamefont [1]{#1}%
\providecommand \citenamefont [1]{#1}%
\providecommand\href[0]{\@sanitize\@href}%
\providecommand\@href[1]{\endgroup\@@startlink{#1}\endgroup\@@href}%
\providecommand\@@href[1]{#1\@@endlink}%
\providecommand \@sanitize [0]{\begingroup\catcode`\&12\catcode`\#12\relax}%
\@ifxundefined \pdfoutput {\@firstoftwo}{%
 \@ifnum{\z@=\pdfoutput}{\@firstoftwo}{\@secondoftwo}%
}{%
 \providecommand\@@startlink[1]{\leavevmode}%
 \providecommand\@@endlink[0]{}%
}{%
 \providecommand\@@startlink[1]{%
  \leavevmode
  \pdfstartlink
   attr{/Border[0 0 1 ]/H/I/C[0 1 1]}%
   user{/Subtype/Link/A<</Type/Action/S/URI/URI(#1)>>}%
  \relax
 }%
 \providecommand\@@endlink[0]{\pdfendlink}%
}%
\providecommand \url  [0]{\begingroup\@sanitize \@url }%
\providecommand \@url [1]{\endgroup\@href {#1}{\urlprefix}}%
\providecommand \urlprefix [0]{URL }%
\providecommand \Eprint[0]{\href }%
\@ifxundefined \urlstyle {%
  \providecommand \doi [1]{doi:\discretionary{}{}{}#1}%
}{%
  \providecommand \doi [0]{doi:\discretionary{}{}{}\begingroup
  \urlstyle{rm}\Url }%
}%
\providecommand \doibase [0]{http://dx.doi.org/}%
\providecommand \Doi[1]{\href{\doibase#1}}%
\providecommand \selectlanguage [0]{\@gobble}%
\providecommand \bibinfo [0]{\@secondoftwo}%
\providecommand \bibfield [0]{\@secondoftwo}%
\providecommand \translation [1]{[#1]}%
\providecommand \BibitemOpen[0]{}%
\providecommand \bibitemStop [0]{}%
\providecommand \bibitemNoStop [0]{.\EOS\space}%
\providecommand \EOS [0]{\spacefactor3000\relax}%
\providecommand \BibitemShut [1]{\csname bibitem#1\endcsname}%
\bibitem{herzberg_book}%
  \BibitemOpen
  \bibfield{author}{%
  \bibinfo {author} {\bibfnamefont{G.}~\bibnamefont{Herzberg}},\ }%
  \emph{\bibinfo {title} {Atomic Spectra and Atomic Structure}}\ (\bibinfo
  {publisher} {Dover Publications},\ \bibinfo {year}
  {1944})\BibitemShut{NoStop}%
\bibitem{meyer_tjocp_1965}%
  \BibitemOpen
  \bibfield{author}{%
  \bibinfo {author} {\bibfnamefont{B.}~\bibnamefont{Meyer}},\ }%
  \bibfield{journal}{%
  \Doi{10.1063/1.1697262}{\bibinfo {journal} {The Journal of Chemical
  Physics}}\ }%
  \textbf{\bibinfo {volume} {43}},\ \bibinfo {pages} {2986} (\bibinfo {year}
  {1965}),\ \url{http://link.aip.org/link/?JCP/43/2986/1}\BibitemShut{NoStop}%
\bibitem{balling_jcp_1978}%
  \BibitemOpen
  \bibfield{author}{%
  \bibinfo {author} {\bibfnamefont{L.~C.}\ \bibnamefont{Balling}}, \bibinfo
  {author} {\bibfnamefont{M.~D.}\ \bibnamefont{Havey}},\ and\ \bibinfo {author}
  {\bibfnamefont{J.~F.}\ \bibnamefont{Dawson}},\ }%
  \bibfield{journal}{%
  \Doi{DOI:10.1063/1.436743}{\bibinfo {journal} {J. Chem. Phys.}}\ }%
  \textbf{\bibinfo {volume} {69}},\ \bibinfo {pages} {1670} (\bibinfo {year}
  {1978}),\ 
  \url{http://link.aip.org/link/doi/10.1063/1.436743}\BibitemShut{NoStop}%
\bibitem{tam_tjocp_1993}%
  \BibitemOpen
  \bibfield{author}{%
  \bibinfo {author} {\bibfnamefont{S.}~\bibnamefont{Tam}}\ and\ \bibinfo
  {author} {\bibfnamefont{M.~E.}\ \bibnamefont{Fajardo}},\ }%
  \bibfield{journal}{%
  \Doi{10.1063/1.465348}{\bibinfo {journal} {The Journal of Chemical Physics}}\
  }%
  \textbf{\bibinfo {volume} {99}},\ \bibinfo {pages} {854} (\bibinfo {year}
  {1993}),\ \url{http://link.aip.org/link/?JCP/99/854/1}\BibitemShut{NoStop}%
\bibitem{crepin-gilbert_iripc_1999}%
  \BibitemOpen
  \bibfield{author}{%
  \bibinfo {author} {\bibfnamefont{C.}~\bibnamefont{Cr\'epin-Gilbert}}\ and\
  \bibinfo {author} {\bibfnamefont{A.}~\bibnamefont{Tramer}},\ }%
  \bibfield{journal}{%
  \bibinfo {journal} {International Reviews in Physical Chemistry}\ }%
  \textbf{\bibinfo {volume} {18}},\ \bibinfo {pages} {485} (\bibinfo {year}
  {1999}),\ 
  \url{http://www.informaworld.com/10.1080/014423599229901}\BibitemShut{NoStop%
}%
\bibitem{moroshkin_pr_2008}%
  \BibitemOpen
  \bibfield{author}{%
  \bibinfo {author} {\bibfnamefont{P.}~\bibnamefont{Moroshkin}}, \bibinfo
  {author} {\bibfnamefont{A.}~\bibnamefont{Hofer}},\ and\ \bibinfo {author}
  {\bibfnamefont{A.}~\bibnamefont{Weis}},\ }%
  \bibfield{journal}{%
  \Doi{DOI: 10.1016/j.physrep.2008.06.004}{\bibinfo {journal} {Physics
  Reports}}\ }%
  \textbf{\bibinfo {volume} {469}},\ \bibinfo {pages} {1 } (\bibinfo {year}
  {2008}),\ 
  \url{http://www.sciencedirect.com/science/article/B6TVP-4T8SKY0-1/2/c2b1d2b6%
2eb255f5790b616eeb7264c7}\BibitemShut{NoStop}%
\bibitem{ryan_tjopca_2010}%
  \BibitemOpen
  \bibfield{author}{%
  \bibinfo {author} {\bibfnamefont{M.}~\bibnamefont{Ryan}}, \bibinfo {author}
  {\bibfnamefont{M.}~\bibnamefont{Collier}}, \bibinfo {author}
  {\bibfnamefont{P.~d.}\ \bibnamefont{Pujo}}, \bibinfo {author}
  {\bibfnamefont{C.}~\bibnamefont{Cr\'epin}},\ and\ \bibinfo {author}
  {\bibfnamefont{J.~G.}\ \bibnamefont{McCaffrey}},\ }%
  \bibfield{journal}{%
  \bibinfo {journal} {The Journal of Physical Chemistry A}\ }%
  \textbf{\bibinfo {volume} {114}},\ \bibinfo {pages} {3011} (\bibinfo {year}
  {2010}),\
  \url{http://pubs.acs.org/doi/abs/10.1021/jp905596a}\BibitemShut{NoStop}%
\bibitem{xu_prl_2011}%
  \BibitemOpen
  \bibfield{author}{%
  \bibinfo {author} {\bibfnamefont{C.-Y.}\ \bibnamefont{Xu}}, \bibinfo {author}
  {\bibfnamefont{S.-M.}\ \bibnamefont{Hu}}, \bibinfo {author}
  {\bibfnamefont{J.}~\bibnamefont{Singh}}, \bibinfo {author}
  {\bibfnamefont{K.}~\bibnamefont{Bailey}}, \bibinfo {author}
  {\bibfnamefont{Z.-T.}\ \bibnamefont{Lu}}, \bibinfo {author}
  {\bibfnamefont{P.}~\bibnamefont{M\"uller}}, \bibinfo {author}
  {\bibfnamefont{T.~P.}\ \bibnamefont{O'Connor}},\ and\ \bibinfo {author}
  {\bibfnamefont{U.}~\bibnamefont{Welp}},\ }%
  \bibfield{journal}{%
  \bibinfo {journal} {Phys Rev Lett}\ }%
  \textbf{\bibinfo {volume} {107}},\ \bibinfo {pages} {093001} (\bibinfo
  {month} {Aug}\ \bibinfo {year} {2011}),\
  \url{http://link.aps.org/doi/10.1103/PhysRevLett.107.093001}\BibitemShut{NoS%
top}%
\bibitem{klempt_pra_2006}%
  \BibitemOpen
  \bibfield{author}{%
  \bibinfo {author} {\bibfnamefont{C.}~\bibnamefont{Klempt}}, \bibinfo {author}
  {\bibfnamefont{T.}~\bibnamefont{van Zoest}}, \bibinfo {author}
  {\bibfnamefont{T.}~\bibnamefont{Henninger}}, \bibinfo {author}
  {\bibfnamefont{O.}~\bibnamefont{Topic}}, \bibinfo {author}
  {\bibfnamefont{E.}~\bibnamefont{Rasel}}, \bibinfo {author}
  {\bibfnamefont{W.}~\bibnamefont{Ertmer}},\ and\ \bibinfo {author}
  {\bibfnamefont{J.}~\bibnamefont{Arlt}},\ }%
  \bibfield{journal}{%
  \Doi{10.1103/PhysRevA.73.013410}{\bibinfo {journal} {Phys. Rev. A}}\ }%
  \textbf{\bibinfo {volume} {73}},\ \bibinfo {pages} {013410} (\bibinfo {month}
  {Jan}\ \bibinfo {year} {2006}),\
  \url{http://link.aps.org/doi/10.1103/PhysRevA.73.013410}\BibitemShut{NoStop}%
\bibitem{kupferman_pr_1968}%
  \BibitemOpen
  \bibfield{author}{%
  \bibinfo {author} {\bibfnamefont{S.~L.}\ \bibnamefont{Kupferman}}\ and\
  \bibinfo {author} {\bibfnamefont{F.~M.}\ \bibnamefont{Pipkin}},\ }%
  \bibfield{journal}{%
  \bibinfo {journal} {Phys. Rev.}\ }%
  \textbf{\bibinfo {volume} {166}},\ \bibinfo {pages} {207} (\bibinfo {month}
  {Feb}\ \bibinfo {year} {1968}),\
  \url{http://dx.doi.org/10.1103/PhysRev.166.207}\BibitemShut{NoStop}%
\bibitem{balling_jcp_1983}%
  \BibitemOpen
  \bibfield{author}{%
  \bibinfo {author} {\bibfnamefont{L.~C.}\ \bibnamefont{Balling}}\ and\
  \bibinfo {author} {\bibfnamefont{J.~J.}\ \bibnamefont{Wright}},\ }%
  \bibfield{journal}{%
  \Doi{DOI:10.1063/1.444487}{\bibinfo {journal} {J. Chem. Phys.}}\ }%
  \textbf{\bibinfo {volume} {78}},\ \bibinfo {pages} {592} (\bibinfo {year}
  {1983}),\ 
  \url{http://dx.doi.org/doi/10.1063/1.444487}\BibitemShut{NoStop}%
\bibitem{eichler_prl_2002}%
  \BibitemOpen
  \bibfield{author}{%
  \bibinfo {author} {\bibfnamefont{T.}~\bibnamefont{Eichler}}, \bibinfo
  {author} {\bibfnamefont{R.}~\bibnamefont{M\"uller-Siebert}}, \bibinfo
  {author} {\bibfnamefont{D.}~\bibnamefont{Nettels}}, \bibinfo {author}
  {\bibfnamefont{S.}~\bibnamefont{Kanorsky}},\ and\ \bibinfo {author}
  {\bibfnamefont{A.}~\bibnamefont{Weis}},\ }%
  \bibfield{journal}{%
  \bibinfo {journal} {Phys. Rev. Lett.}\ }%
  \textbf{\bibinfo {volume} {88}},\ \bibinfo {pages} {123002} (\bibinfo {month}
  {Mar}\ \bibinfo {year} {2002}),\
  \url{http://dx.doi.org/10.1103/PhysRevLett.88.123002}\BibitemShut{NoStop}%
\bibitem{hofer_pra_2006}%
  \BibitemOpen
  \bibfield{author}{%
  \bibinfo {author} {\bibfnamefont{A.}~\bibnamefont{Hofer}}, \bibinfo {author}
  {\bibfnamefont{P.}~\bibnamefont{Moroshkin}}, \bibinfo {author}
  {\bibfnamefont{D.}~\bibnamefont{Nettels}}, \bibinfo {author}
  {\bibfnamefont{S.}~\bibnamefont{Ulzega}},\ and\ \bibinfo {author}
  {\bibfnamefont{A.}~\bibnamefont{Weis}},\ }%
  \bibfield{journal}{%
  \Doi{10.1103/PhysRevA.74.032509}{\bibinfo {journal} {Phys. Rev. A}}\ }%
  \textbf{\bibinfo {volume} {74}},\ \bibinfo {pages} {032509} (\bibinfo {month}
  {Sep}\ \bibinfo {year} {2006}),\
  \url{http://link.aps.org/doi/10.1103/PhysRevA.74.032509}\BibitemShut{NoStop}%
\bibitem{kinoshita_prb_1994}%
  \BibitemOpen
  \bibfield{author}{%
  \bibinfo {author} {\bibfnamefont{T.}~\bibnamefont{Kinoshita}}, \bibinfo
  {author} {\bibfnamefont{Y.}~\bibnamefont{Takahashi}},\ and\ \bibinfo {author}
  {\bibfnamefont{T.}~\bibnamefont{Yabuzaki}},\ }%
  \bibfield{journal}{%
  \bibinfo {journal} {Phys. Rev. B}\ }%
  \textbf{\bibinfo {volume} {49}},\ \bibinfo {pages} {3648} (\bibinfo {month}
  {Feb}\ \bibinfo {year} {1994}),\
  \url{http://link.aps.org/doi/10.1103/PhysRevB.49.3648}\BibitemShut{NoStop}%
\bibitem{kinoshita_pra_1995}%
  \BibitemOpen
  \bibfield{author}{%
  \bibinfo {author} {\bibfnamefont{T.}~\bibnamefont{Kinoshita}}, \bibinfo
  {author} {\bibfnamefont{K.}~\bibnamefont{Fukuda}}, \bibinfo {author}
  {\bibfnamefont{Y.}~\bibnamefont{Takahashi}},\ and\ \bibinfo {author}
  {\bibfnamefont{T.}~\bibnamefont{Yabuzaki}},\ }%
  \bibfield{journal}{%
  \bibinfo {journal} {Phys. Rev. A}\ }%
  \textbf{\bibinfo {volume} {52}},\ \bibinfo {pages} {2707} (\bibinfo {month}
  {Oct}\ \bibinfo {year} {1995}),\
  \url{http://link.aps.org/doi/10.1103/PhysRevA.52.2707}\BibitemShut{NoStop}%
\bibitem{theisen_epjd_2011}%
  \BibitemOpen
  \bibfield{author}{%
  \bibinfo {author} {\bibfnamefont{M.}~\bibnamefont{Theisen}}, \bibinfo
  {author} {\bibfnamefont{F.}~\bibnamefont{Lackner}}, \bibinfo {author}
  {\bibfnamefont{F.}~\bibnamefont{Ancilotto}}, \bibinfo {author}
  {\bibfnamefont{C.}~\bibnamefont{Callegari}},\ and\ \bibinfo {author}
  {\bibfnamefont{W.}~\bibnamefont{Ernst}},\ }%
  \bibfield{journal}{%
  \Doi{10.1140/epjd/e2010-10504-5}{\bibinfo {journal} {Eur. Phys. J. D}}\ }%
  \textbf{\bibinfo {volume} {61}},\ \bibinfo {pages} {403} (\bibinfo {year}
  {2011}),\
  \url{http://dx.doi.org/10.1140/epjd/e2010-10504-5}\BibitemShut{NoStop}%
\bibitem{micklitz_zfpahan_1974}%
  \BibitemOpen
  \bibfield{author}{%
  \bibinfo {author} {\bibfnamefont{H.}~\bibnamefont{Micklitz}}\ and\ \bibinfo
  {author} {\bibfnamefont{K.}~\bibnamefont{Luchner}},\ }%
  \bibfield{journal}{%
  \bibinfo {journal} {Zeitschrift fur Physik A Hadrons and Nuclei}\ }%
  \textbf{\bibinfo {volume} {270}},\ \bibinfo {pages} {79} (\bibinfo {month}
  {Mar.}\ \bibinfo {year} {1974}),\
  \url{http://dx.doi.org/10.1007/BF01676798}\BibitemShut{NoStop}%
\bibitem{smith_pm_1961}%
  \BibitemOpen
  \bibfield{author}{%
  \bibinfo {author} {\bibfnamefont{B.~L.}\ \bibnamefont{Smith}},\ }%
  \bibfield{journal}{%
  \bibinfo {journal} {Philosophical Magazine}\ }%
  \textbf{\bibinfo {volume} {6}},\ \bibinfo {pages} {939} (\bibinfo {year}
  {1961}),\
  \url{http://dx.doi.org/10.1080/14786436108243350}\BibitemShut{NoStop}%
\bibitem{sinnock_pr_1969}%
  \BibitemOpen
  \bibfield{author}{%
  \bibinfo {author} {\bibfnamefont{A.~C.}\ \bibnamefont{Sinnock}}\ and\
  \bibinfo {author} {\bibfnamefont{B.~L.}\ \bibnamefont{Smith}},\ }%
  \bibfield{journal}{%
  \Doi{10.1103/PhysRev.181.1297}{\bibinfo {journal} {Phys. Rev.}}\ }%
  \textbf{\bibinfo {volume} {181}},\ \bibinfo {pages} {1297} (\bibinfo {month}
  {May}\ \bibinfo {year} {1969}),\
  \url{http://link.aps.org/doi/10.1103/PhysRev.181.1297}\BibitemShut{NoStop}%
\bibitem{sinnock_jopcssp_1980}%
  \BibitemOpen
  \bibfield{author}{%
  \bibinfo {author} {\bibfnamefont{A.~C.}\ \bibnamefont{Sinnock}},\ }%
  \bibfield{journal}{%
  \bibinfo {journal} {Journal of Physics C: Solid State Physics}\ }%
  \textbf{\bibinfo {volume} {13}},\ \bibinfo {pages} {2375} (\bibinfo {year}
  {1980}),\
  \url{http://stacks.iop.org/0022-3719/13/i=12/a=018}\BibitemShut{NoStop}%
\bibitem{bondi_tjopc_1964}%
  \BibitemOpen
  \bibfield{author}{%
  \bibinfo {author} {\bibfnamefont{A.}~\bibnamefont{Bondi}},\ }%
  \bibfield{journal}{%
  \Doi{10.1021/j100785a001}{\bibinfo {journal} {The Journal of Physical
  Chemistry}}\ }%
  \textbf{\bibinfo {volume} {68}},\ \bibinfo {pages} {441} (\bibinfo {year}
  {1964}),\
  \url{http://pubs.acs.org/doi/abs/10.1021/j100785a001}\BibitemShut{NoStop}%
\bibitem{mantina_tjopca_2009}%
  \BibitemOpen
  \bibfield{author}{%
  \bibinfo {author} {\bibfnamefont{M.}~\bibnamefont{Mantina}}, \bibinfo
  {author} {\bibfnamefont{A.~C.}\ \bibnamefont{Chamberlin}}, \bibinfo {author}
  {\bibfnamefont{R.}~\bibnamefont{Valero}}, \bibinfo {author}
  {\bibfnamefont{C.~J.}\ \bibnamefont{Cramer}},\ and\ \bibinfo {author}
  {\bibfnamefont{D.~G.}\ \bibnamefont{Truhlar}},\ }%
  \bibfield{journal}{%
  \Doi{10.1021/jp8111556}{\bibinfo {journal} {The Journal of Physical Chemistry
  A}}\ }%
  \textbf{\bibinfo {volume} {113}},\ \bibinfo {pages} {5806} (\bibinfo {year}
  {2009}),\
  \url{http://pubs.acs.org/doi/abs/10.1021/jp8111556}\BibitemShut{NoStop}%
\bibitem{crc_pola}%
  \BibitemOpen
  \emph{\bibinfo {title} {Handbook of Chemistry and Physics}},\ \bibinfo
  {edition} {91st}\ ed.\ (\bibinfo {publisher} {CRC},\ \bibinfo {year}
  {2011})\BibitemShut{NoStop}%
\bibitem{schrimpf_zfpbcm_1987}%
  \BibitemOpen
  \bibfield{author}{%
  \bibinfo {author} {\bibfnamefont{A.}~\bibnamefont{Schrimpf}}, \bibinfo
  {author} {\bibfnamefont{G.}~\bibnamefont{Sulzer}}, \bibinfo {author}
  {\bibfnamefont{H.~J.}\ \bibnamefont{St\"{o}ckmann}},\ and\ \bibinfo {author}
  {\bibfnamefont{H.}~\bibnamefont{Ackermann}},\ }%
  \bibfield{journal}{%
  \bibinfo {journal} {Zeitschrift f\"{u}r Physik B Condensed Matter}\ }%
  \textbf{\bibinfo {volume} {67}},\ \bibinfo {pages} {531} (\bibinfo {year}
  {1987}),\
  \url{http://dx.doi.org/10.1007/BF01304125}\BibitemShut{NoStop}%
\end{thebibliography}

%
\end{document}